\documentclass[12pt,preprint]{aastex}

\usepackage{psfig}

\begin{document}
\title{High-Resolution Spectroscopy of G191-B2B in the Extreme Ultraviolet}
\shorttitle{High Resolution EUV Spectrum of G191-B2B}
\author{R. G. Cruddace\altaffilmark{1}, M . P. Kowalski\altaffilmark{1},
D. Yentis\altaffilmark{1}, C. M. Brown\altaffilmark{1}, 
H. Gursky\altaffilmark{1}}
\author{M. A. Barstow\altaffilmark{2}, N. P. Bannister\altaffilmark{2},
G. W. Fraser\altaffilmark{2}, J. E. Spragg\altaffilmark{2}}
\author{J. S. Lapington\altaffilmark{3}, J. A. Tandy\altaffilmark{3},
B. Sanderson\altaffilmark{3}, J. L. Culhane\altaffilmark{3}}
\author{T. W. Barbee\altaffilmark{4}, J. F. Kordas\altaffilmark{4},
W. Goldstein\altaffilmark{4}}
\author{and G. G. Fritz\altaffilmark{5}}
\altaffiltext{1}{E. O. Hulburt Center for Space Research, Naval
Research Laboratory, Washington DC 20375, U.S.A.}
\altaffiltext{2}{Department of Physics and Astronomy, University of Leicester,
Leicester LE1 7RH, U.K.}
\altaffiltext{3}{Mullard Space Science Laboratory, University College London,
Holmbury St. Mary, Dorking, Surrey RH5 6NT, U.K.}
\altaffiltext{4}{Lawrence Livermore National Laboratory, Livermore, California 
97550, U.S.A.}
\altaffiltext{5}{Praxis Inc., 2200 Mill Road, Alexandria, Virginia 22314, 
U.S.A.}

\begin{abstract}
We report a high-resolution (R=3000-4000) spectroscopic 
observation of the DA white dwarf G191-B2B in the extreme ultraviolet band 
$220-245$\ \AA . A low-density ionised He component is clearly present along the
line-of-sight, which if completely interstellar implies a He ionisation fraction considerably higher than is 
typical of the local interstellar medium. However, some of this material may
be associated with circumstellar gas, which has been detected by analysis of
the CIV absorption line doublet in an HST STIS spectrum. A stellar atmosphere 
model assuming a uniform element distribution yields a best fit to the data 
which includes a significant abundance of photospheric He. The 99-percent 
confidence contour for the fit parameters excludes solutions in which 
photospheric He is absent, but this result needs to be tested using 
models allowing abundance gradients.
\end{abstract}

\keywords{white dwarfs---stars: individual(G191B2B)---ISM: general---
circumstellar matter}

\section*{1.  INTRODUCTION}
White dwarfs are among the oldest objects in the Galaxy. As remnants of all
stars with an initial mass $<\,8\,{\rm M}_{\odot }$, they are
important laboratories for study of evolutionary processes and the behavior
of matter at extreme temperature and density. Study of their space and 
luminosity distributions helps map the history of star formation and could, 
in principle, determine the age of the disk, yielding an important lower 
limit to the age of the Universe. Further, cool 
white dwarfs may account for a substantial fraction of the missing mass in the 
galactic halo \citep{opp01}. However, these goals depend on
our understanding of white dwarf evolution, and in particular on predictions 
of the cooling rates. These in turn are affected by the mass, radius and 
photospheric composition of the stars.
Our understanding of the physical mechanisms that determine white dwarf 
evolution leaves unanswered several major questions.  While the emergence from 
the AGB of two groups of white dwarfs, whose compositions are dominated by H or
He, is beginning to be understood, the complex relationship between these 
branches, and a demonstrable temperature gap in the cooling sequence of the
He-rich branch, cannot yet be explained. Determination of the
photospheric He and heavy element content would provide important information 
on the evolutionary history of the stars. Already it has been established 
that significant quantities of elements heavier than He are present in the
atmospheres of the hottest (${\rm T}>50,000\,{\rm K}$) white dwarfs 
\citep{mab98}. 

G191-B2B is one of the brightest and best-studied of the hot H-rich DA white 
dwarfs. As it lies near the top of the DA cooling sequence,
measurements of its effective temperature, surface gravity and composition
represent an important benchmark in the study of the whole DA sample. 
IUE and HST observations at high spectral resolution have made it clear,
that G191-B2B falls into the group of very hot DA stars 
(${\rm T}>50,000\,{\rm K}$) whose atmospheres contain significant quantities of
heavy elements. In particular, detections of C, N, O, Si, P, S, Fe and Ni 
have been reported in various studies \citep{bk83,si92,ven92,ven96,jh94}.
Such material is 
responsible for severe depression of the extreme ultraviolet (EUV) flux in 
G191-B2B at $\lambda \,<\,200$\ \AA , when it is compared with 
stars having pure H atmospheres, and the star has been an important target 
for spectroscopic EUV observations to determine the principal opacity sources. 
In addition, an important goal is to obtain a self-consistent model, having an 
effective temperature, surface gravity and composition which can fit the 
far ultraviolet (FUV) and EUV observations simultaneously. This would 
demonstrate our understanding of the star and the reliability of 
the model calculations, which could then be applied to other objects.

However, a complete understanding of the EUVE spectrum of G191-B2B has been 
elusive. Initial attempts to match the observation with synthetic spectra
failed to reproduce either the flux level or the general shape of the
continuum (see \citet{mab96}), as the model contained
insufficient Fe and Ni lines. Consequently $\sim 9$ million predicted lines were
added to the few thousand with measured wavelengths, yielding a self-consistent
model able to reproduce the EUV, FUV and optical spectra \citep{l96}. However, 
good agreement could be achieved only by 
including a significant quantity of He, either in the photosphere 
or in an ionised interstellar component. Unfortunately, due to 
the limited resolution of EUVE ($\sim 0.5$\ \AA ) in the  
He II Lyman series band, the He contribution could not be detected directly.
More recently, it has been shown that photospheric heavy elements may not be
distributed homogeneously in the radial direction, so that more complex 
stratified structures should be considered. This approach has yielded a good 
fit of model atmospheres to observational data across the soft X-ray, EUV and FUV 
bands \citep{bhh99,dw99}.
Important progress has been made also in incorporating radiative 
levitation and diffusion self-consistently into the calculations 
\citep{dw99,sdw01}. The need for a He contribution is 
reduced in these stratified models but not eliminated. We present the 
results of a search for the He II component in 
the EUV spectrum of G191-B2B, using the J-PEX (Joint Astrophysical 
Plasmadynamic Experiment) high-resolution EUV spectrometer. This spectrum 
was obtained by J-PEX after launch by a sounding rocket in February 2001.
\section*{2.  The J-PEX High Resolution EUV Spectrometer}
The J-PEX spectrometer, described by Cruddace et al. (in preparation),
is a slitless, normal-incidence instrument employing figured spherical 
gratings in a Wadsworth mount. 
The optic comprises four ion-etched laminar gratings with a groove density
of $3600\,{\rm g}\,{\rm mm}^{-1}$ and a focal length of $2.0\,{\rm m}$. The
gratings are coated with a ${\rm Mo}_{5}{\rm C}/{\rm Si}/{\rm MoSi}_{2}$ 
multilayer designed to operate in the
band $220-245$\ \AA , and are optimised to suppress
zero order and to yield maximum efficiency in first order at 
$235$\ \AA . At this wavelength the spectrometer achieves an effective area of
$3.0\,{\rm cm}^{2}$. The design specification for the spectral resolving power 
(R) in flight is $4980$, which includes the effect of a pointing uncertainty of 
$1\,{\rm arcsec}$. Calibration of R was hindered by thermal deflections in the
spectrometer, caused by detector heating over long exposure times, and 
consequenctly the average measured resolving power
was 2750, yielding an estimate for inflight resolving power of 2600. However
the problem encountered in calibration should have a much smaller effect 
during a flight lasting only 5 minutes. Therefore we can only
set reasonable bounds to the resolving power in flight, namely
$3000<{\rm R}<4000$. This uncertainty is taken into account in analysis of the
flight data. J-PEX was launched by a NASA Black Brant IX sounding rocket 
(NASA 36.195DG)
at 05.45UT on 22 February 2001. The payload completed its mission successfully,
during which it observed the target G191-B2B for $300$ seconds.
\section*{3.  Data Reduction and Analysis}
\begin{figure*}
\centerline{\psfig{figure=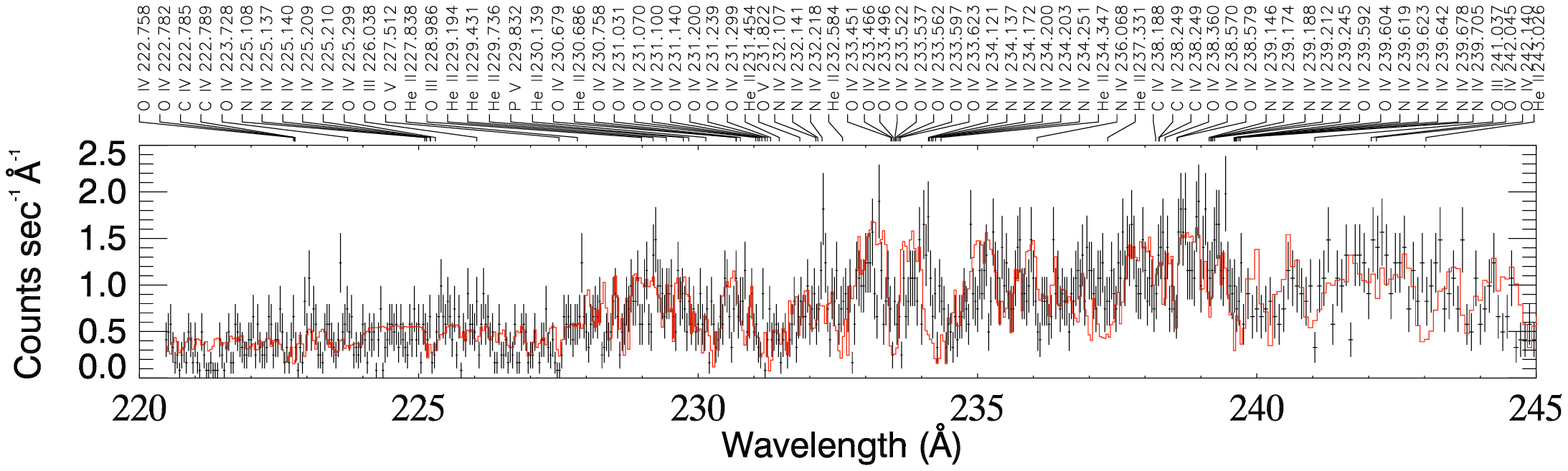,width=6.5in,angle=90}}
\caption{High resolution EUV spectrum of G191-B2B, obtained with the J-PEX 
spectrometer in the waveband $221-244$\ \AA . All data points 
have error bars. The red histogram is the best-fit model of the star and ISM. 
The strongest predicted lines of He, C, N, O, and P are labeled 
with their ionization state and wavelength. Lines of Fe and Ni, too numerous
to include here, account for some unlabelled individual features and 
broader absorption structures.}
\label{f1}
\end{figure*}

The spectra of the four gratings were recorded independently by the focal-plane
detector. In Cruddace et al. (in preparation) we describe how the positions of 
photon events in each spectrum were corrected for ACS drift and 
jitter, and the wavelength of each spectrum was calibrated.
The events were summed in bins of width $0.024$\ \AA , before
being superposed to yield one spectrum. The bin width was chosen to 
oversample the data by a factor 2.4 in comparison with the resolution
at ${\rm R}=4000$, so as to minimise loss of information during
superposition. The final spectrum, in which the signal-to-noise ratio (S/N) 
has been maximised by increasing the bin size to $0.048$\ \AA , is shown in 
Figure \ref{f1}. The average S/N per bin was 5.0. 
The background count-rate in the detector field during the observation was 
$4.2\,{\rm ct}\,{\rm s}^{-1}$, sufficiently low that the spectra were 
essentially free of background. Therefore the error bars in Figure \ref{f1}
have been assigned assuming Poisson statistics. The spectrum contains one flaw
caused by the pointing on target, in which data above $\sim 239$\ \AA \ was 
lost in two of the spectra. Thus only half the instrument effective area was 
used in this region, in which accordingly we have increased the bin width to
$0.096$\ \AA . 

We have compared the observed spectrum with the predictions of a model,
in which the white dwarf atmosphere has a homogeneous composition and the flux
is absorbed by H I, He I and He II in the interstellar medium (ISM). 
Although stratified atmosphere 
models are more successful in reconciling spectra of G191-B2B in the EUV and FUV
bands \citep{bh98,bhh99}, the homogeneous models are adequate 
for the relatively narrow J-PEX waveband, and give a useful baseline for 
comparison with abundance measurements made in other studies. The analysis 
technique has been described extensively in 
earlier papers (e.g. Lanz et al. 1996), and we give here only a 
brief overview. The XSPEC software was used to fold model spectra through the 
J-PEX instrument response, and as we were dealing with a spectrum having a 
small number of counts per bin, the best match between model 
and data was obtained by minimisation of the Cash statistic \citep{wc79}.
This does not assign an absolute value to the goodness of fit, but does 
allow uncertainty ranges to be determined for each free parameter.

The model spectra, based on work reported by \citet{l96} and 
\citet{bhh98,bhh99}, were calculated using the non-LTE code 
TLUSTY \citep{hl95}. For this initial analysis, we 
fixed the stellar temperature and surface gravity 
(${\rm T}_{\rm eff}=54,000\,{\rm K}$, ${\rm log\,g}=7.5$) at the grid points 
closest to the values determined using
the Balmer and Lyman lines \citep{bhh98}.
Apart from the He abundance, which was allowed to vary freely between the grid 
limits of 
$10^{-4}$ and $10^{-6}$, the heavy element abundances were fixed at values
determined in earlier homogeneous-model analyses of G191-B2B  
(${\rm C/H}=4.0\times 10^{-7}$, ${\rm N/H}=1.\times 10^{-7}$, 
${\rm O/H}=9.6\times 10^{-7}$, ${\rm Si/H}=3.0\times 10^{-7}$,
${\rm P/H}=2.5\times 10^{-8}$, ${\rm S/H}=3.2\times 10^{-7}$,
${\rm Fe/H}=1.0\times 10^{-5}$, ${\rm Ni/H}=5.0\times 10^{-7}$). 

\begin{figure*}
\centerline{\psfig{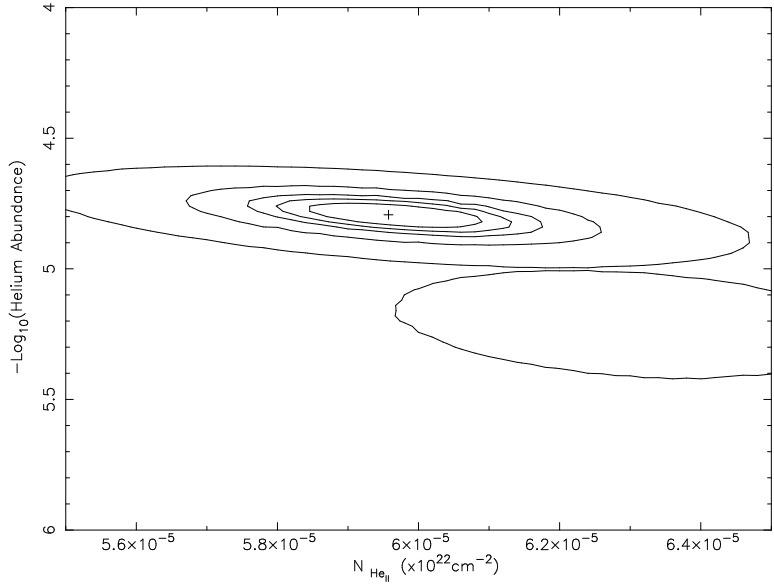}}
\caption{The 1, 2, 3, 5 and 10 sigma contours for the two parameters varied in
fitting a uniform abundance white dwarf atmosphere model to the measured
spectrum of G191-B2B. The 3 sigma contour corresponds to a confidence level of 
99.7 percent. The secondary contour at 10 sigma locates a weak secondary minimum
in the $\chi ^{2}$ distribution.}
\label{f2}
\end{figure*}

The value taken for Fe/H lies between limits established by FUV
($2.4\times 10^{-6}$, \citet{ven01}) and EUV ($3-4\times 10^{-5}$, \citet{bhh99}) 
analyses. The
effect of Fe/H in the $225-245$\ \AA \ band is to change the level of the 
overall
spectrum, and we have verified that this does not affect the conclusions 
reached in our analysis. The ISM H I and He I column densities were 
fixed at values obtained from analysis of the broader-band, lower-resolution 
EUVE spectrum \citep{bhh99}: 
${\rm H\ I}=2.15\times 10^{18}\,{\rm cm}^{-2}$, 
${\rm He\ I}=2.18\times 10^{17}\,{\rm cm}^{-2}$. The parameters varied during
the fit were the column density of He II in the line-of-sight 
(${\rm N}_{\rm He II}$) and the
photospheric He abundance (${\rm n}_{\rm He}$: measured by numbers of nuclei). 
For this fit, the data were summed in bins of width $0.060$\ \AA , equivalent to 
${\rm R}=4000$. Given the uncertainty in the resolving power 
(section 2, ${\rm R}=3000-4000$), the fits were performed also for lower values
of R, but yielded no significant change in the results presented below in
section 4. The best fit to the data, shown by the red line in Figure \ref{f1},
was obtained for ${\rm N}_{\rm He II}=5.97\times 10^{17}\,{\rm cm}^{-2}$ and 
${\rm n}_{\rm He}=1.60\times 10^{-5}$, and in Figure \ref{f2} 
we show the 1, 2, 3, 5 and 10 sigma
contours for the two parameters. The 3 sigma contour is the locus on
which $\chi^{2}$ exceeds
the minimum by $11.8$, and within which the parameter confidence level is
greater than $99.7$ percent. We use this contour to derive 99-percent confidence
limits of $5.76-6.18\times 10^{17}\,{\rm cm}^{-2}$ for 
${\rm N}_{\rm He II}$ and $1.31-1.91\times 10^{-5}$ for 
${\rm n}_{\rm He}$.
\section*{4.  Discussion}
\begin{figure*}
\centerline{\psfig{figure=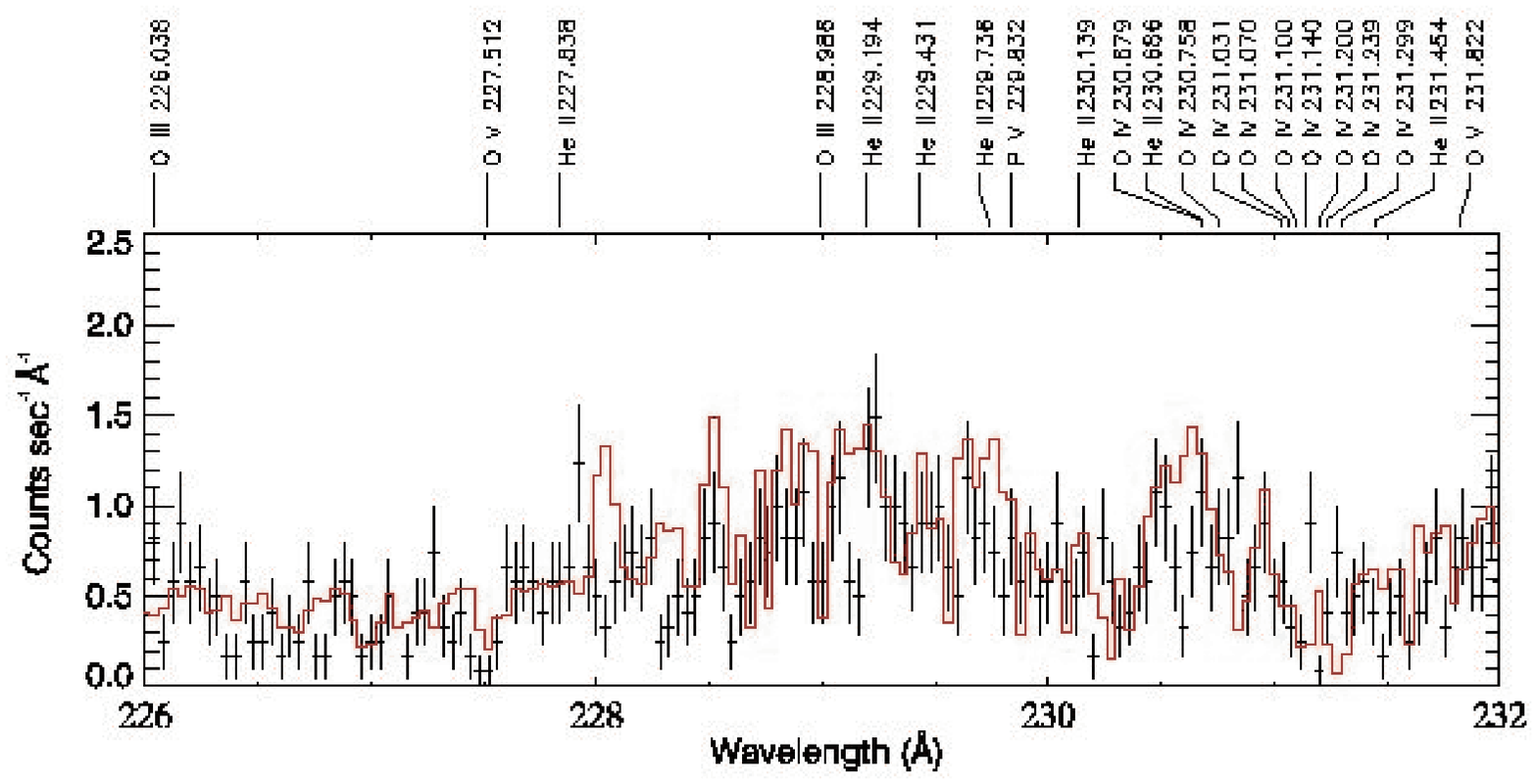,width=4.2in,angle=0}}
\centerline{\psfig{figure=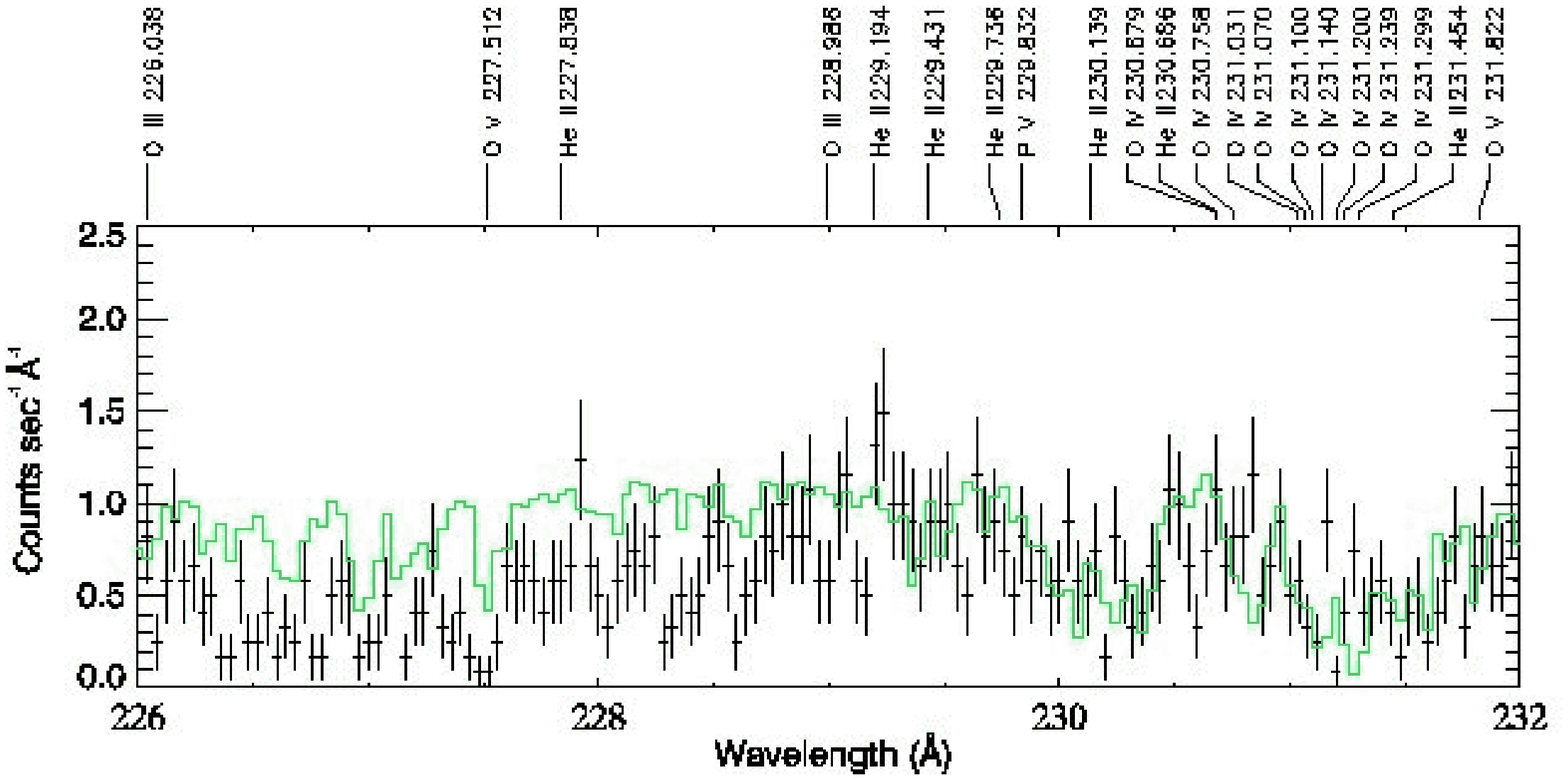,width=4.2in,angle=0}}
\caption{Upper panel: Expanded view of the J-PEX spectrum of G191-B2B 
in the wavelength range $226-232$ \AA \, spanning the He II Lyman series limit. 
The red histogram is the best-fitting model of the stellar atmosphere and 
absorption in the ISM.  
\newline Lower panel: Expanded view of the J-PEX spectrum of G191-B2B in the 
range $226-232$ \AA \, but this time showing the best fit, traced by the green 
histogram, of a model in which 
absorption by interstellar and circumstellar He has been removed.}
\label{f3}
\end{figure*}

The good agreement between the best-fit model and the data in Figure \ref{f1}
is striking, for example at the prominent absorption feature at  
$233.5$\ \AA \ produced by a cluster of O IV lines. Many other 
features are present, which are mainly blends of large numbers of Fe V and Ni V 
lines. The broad features between 227 and 232 \AA \ are a characteristic of the
overlapping series of interstellar He II absorption lines superposed on a
continuum. This is shown more clearly in the upper panel of Figure \ref{f3}, 
an expanded view of the region $226-232$ \AA \ in Figure \ref{f1}. 
Taken with the
strong depression of the flux below $227$ \AA \, this is strong evidence that 
interstellar He II is present along the line of sight. 
Conclusive proof is obtained when the data are fitted by a model in which
${\rm N}_{\rm He II}$ is set to zero. The degradation of the fit is
evident in the lower panel of Figure \ref{f3}, particularly in the region below
$229$\ \AA \ . Further, in this case the best-fit value of $8.0\times 10^{-5}$ 
for ${\rm n}_{\rm He}$ is about four times the upper limit we
obtained from the absence of detectable He II at $1640$ \AA \ in the STIS 
spectrum of G191-B2B. On the other hand, the best-fit model in
the upper panel of Figure \ref{f3}, in which ${\rm n}_{\rm He}=1.6\,10^{-5}$, 
is consistent with the STIS limit.

The best-fit He II column density of $5.97\times 10^{17}\,{\rm cm}^{-2}$
implies a He ionization fraction, based on EUVE measurements of the He I 
column density, of $\sim 0.73$. This is substantially higher than the typical
range of 0.25-0.50 in the local interstellar medium (LISM; e.g. Barstow 
et al. 1997). However, a possible circumstellar (CSM) component has been
identified recently through analysis of the CIV absorption line doublet
($1548.202\,{\rm and}\,1550.774$\ \AA ) in the STIS FUV spectrum of G191-B2B 
\citep{npb01}. Therefore some of the He II
detected in the J-PEX spectrum may be circumstellar, although the fraction
would have to be at least one third to bring the interstellar component within
the LISM range. The known LISM and CSM components are separated by 
$\sim 8\,{\rm km}\,{\rm s}^{-1}$, and therefore could not be resolved by J-PEX.

The photospheric absorption line at $243.026$ \AA \ is predicted to be the 
strongest in the J-PEX waveband. Unfortunately this lies in a region
of reduced exposure (Section 3), and Figure \ref{f1} shows no feature 
in the measured spectrum at this wavelength. However, the depth of the predicted
line is similar to the
magnitude of the statistical errors for the data points, yielding only a very
weak constraint on the possible He abundance. A weak absorption
line is seen at the position of He II 237.331 \AA \, but likewise the photon 
statistics do not allow a significant detection.

The above discussion of the column density of ISM/CSM
He II, and the evidence for photospheric He, is underpinned by the good 
agreement between the homogeneous stellar atmosphere models and the 
observational data, as shown by the good correspondence between 
predicted and observed features in Figure \ref{f1} and the upper panel
of Figure \ref{f3}. However, closer inspection also reveals several 
significant features present in the observed but not in the 
synthetic spectrum, for example at $229.3$ and $231.2$ \AA \ . This
indicates that the model atmosphere is incomplete 
and that other elements should be included in addition to C, N, O, Si, P, S, Fe
and Ni. 
\section*{5.  Conclusions}
We have presented our first analysis of the high-resolution EUV spectrum of
G191-B2B obtained with the J-PEX spectrometer, which has 
the highest resolving power (3000-4000) achieved so far in the EUV and X-ray 
wavebands. The results show conclusively that ionised He is present along
the line-of-sight, and yield a column density of 
$5.97\times 10^{17}\,{\rm cm}^{-2}$. However if this is all 
interstellar, the measured column density yields an estimated He ionisation 
fraction  significantly higher than is typical of the LISM. Some of 
this material may be associated with a newly discovered circumstellar component,
but further 
investigation is necessary to demonstrate whether this is plausible.

The white dwarf model which best fitted the data included a 
significant abundance, $1.6\times 10^{-5}$, of
photospheric He, and at the 99-percent confidence level we could
exclude models containing no photospheric He. However, we caution
that a simple model, in which elements are
distributed uniformly in the atmosphere, was used in this analysis, and
further work with models in which stratification of elements is allowed,
and taking greater advantage of other G191-B2B observations, is needed to
verify or disprove our result.

\acknowledgements
The Naval Research Laboratory was supported in this work by NASA
under the grant NDPR S-47440F, and by the Office of Naval Research under NRL
Work Unit 3641 (Application of Multilayer Coated Optics to Remote Sensing). 
The University of Leicester and the Mullard Space Science Laboratory 
acknowledge the support they received for this project from the Particle 
Physics and Astronomy Research Council in the U. K.

\end{document}